\documentclass[pdflatex,sn-nature]{sn-jnl}

\usepackage{graphicx}%
\usepackage{multirow}%
\usepackage{amsmath,amssymb,amsfonts}%
\usepackage{amsthm}%
\usepackage{mathrsfs}%
\usepackage[title]{appendix}%
\usepackage{xcolor}%
\usepackage{textcomp}%
\usepackage{manyfoot}%
\usepackage{booktabs}%
\usepackage{algorithm}%
\usepackage{algorithmicx}%
\usepackage{algpseudocode}%
\usepackage{listings}%
\usepackage{mdframed}
\usepackage{caption}
\usepackage{newfloat}
\usepackage{placeins}

\DeclareCaptionLabelFormat{supplement_figure_label_fmt}{\textbf{Extended Data Fig.~#2}}

\newenvironment{supplementfigure}
  {\captionsetup{labelformat=supplement_figure_label_fmt,labelsep=colon}
   \begin{figure}}
  {\end{figure}}

\theoremstyle{thmstyleone}%
\theoremstyle{thmstyletwo}%

\theoremstyle{thmstylethree}%

\newcommand{\methodhead}[1]{\par\vspace{1.5ex}\noindent\textbf{\textsc{#1}}.\ }

\begin{document}


\title[Article Title]{Asymmetric Impact of Basic Scientists during Applied Shift}


\author[1]{\fnm{Rikuei} \sur{Kaku}}

\author[1]{\fnm{Mikako} \sur{Bito}}

\author[1]{\fnm{Keita} \sur{Nishimoto}}

\author[1]{\fnm{Ichiro} \sur{Sakata}}

\author*[1]{\fnm{Kimitaka} \sur{Asatani}}\email{asatani@tmi.t.u-tokyo.ac.jp}

\affil*[1]{\orgdiv{Graduate School of Engineering}, \orgname{The University of Tokyo}, \orgaddress{\street{7-3-1 Hongo}, \city{Bunkyo City}, \postcode{113-8654}, \state{Tokyo}, \country{Japan}}}


\abstract{
Despite broad acclaim for basic research, science is undergoing an applied shift that marginalizes basic scientists. This gap reflects an incomplete understanding of their distinctive roles, which prevents translating philosophical appreciation into effective support. We introduce a scalable metric—the application score—to position research along the basic–applied spectrum and apply it to 62 million publications (1970–2023) to reveal the distinctive contributions of basic scientists. We find a structural asymmetry: involvement of basic scientists substantially increases citation impact, even more so in applied contexts, while applied scientists show no such effect in basic domains. This asymmetric effect arises from their intellectual leadership in conceptualization, writing, and experimental design, amplified in large, multidisciplinary, and intermediate career teams. Yet basic scientists remain concentrated in historically prestigious institutions, while new entrants shift toward applied work, indicating critical undersupply. These findings provide large-scale evidence for the indispensable role of basic scientists, guiding policy and institutional strategy to sustain the foundations of discovery and innovation.
}

\maketitle


\section{Introduction}\label{sec:introduction}

Basic research drives scientific progress\cite{bushEndless1945,schauzWhat2014}, essential for both long-term discovery\cite{nelson1959simple,salterEconomic2001} and near-term innovation\cite{ahmadpoor2017dual,narin1997increasing}. Its value has traditionally been recognized through the production of high-impact, foundational papers within individual domains\cite{martinAssessing1983}. Basic scientists shape more than theoretical foundations, as their participation enhances outcomes across the research spectrum, from fundamental discoveries to practical applications. Classic and contemporary examples include Pasteur's microbiology contributing to advances in public health\cite{donald1997stokes} and, more recently, Hassabis' vision, shaped by his background in cognitive neuroscience, and Jumper’s computational foundations, which enabled biomedical breakthroughs through AlphaFold\cite{jumper2021highly}. These cases reveal how basic scientists generate scientific impact in foundational and application-oriented domains alike.

Despite such celebrated achievements, basic scientists face growing challenges\cite{Funk_Rainie_2015}. The numbers tell the story: in the United States, the federal government's share of basic research funding has dropped from over 70\% to around 40\%\cite{aaasDataCheck2018}, and by 2021 basic research accounted for only 15\% of total R\&D expenditure—a historic low\cite{nsfIndicators2024}. Universities, traditionally centers of fundamental inquiry, now prioritize applied agendas, emphasizing translational impact and short-term returns. Career paths reflect this shift: new Ph.D.s are nearly as likely to enter the private sector (42\%) as academia (43\%)\cite{ganapati2021professional}, and even academic researchers encounter constant pressure to orient their work toward applied priorities for funding and stability\cite{bentley2015relationship,sauermannScience2012}. Paradoxically, while institutions drift away from basic science, basic scientists’ expertise is more valued than ever~\cite{stephan2012economics}. This creates a fundamental contradiction: the very actors who provide the intellectual foundation for applied advances find themselves in increasingly unstable positions.

Quantifying the scientific and societal impact of basic scientists is crucial, providing evidence for fundamental inquiry's value beyond philosophical arguments. Yet defining what counts as ``basic'' remains elusive. As P.~W.\ Anderson argued, ``more is different'': new fundamental principles emerge at higher levels of complexity, implying that the notion of ``basicness'' varies across domains and resists a universal checklist~\cite{anderson1972more}. Measuring how basic a given research output is, therefore, poses a persistent methodological challenge. Citation-based proxies capture certain aspects: fundamental work tends to diffuse broadly across disciplines~\cite{wangQuantifying2013}, and patents disproportionately cite upstream science~\cite{ahmadpoor2017dual}. More recent approaches use text-based methods and domain-specific embeddings to position research along the basic–applied continuum~\cite{keIdentifying2019,higashideQuantifying2024}. Meanwhile, large language models (LLMs) offer potential for emulating expert judgment, supporting evaluation, and even acting as collaborators in science~\cite{birhane2023science, zhang2025evolving}.

Here, we develop a scalable, cross-disciplinary metric using machine learning and LLMs to score research on a basic–applied spectrum, based on Bush's linear model \cite{bushEndless1945}.  Analyzing 62 million publications from 1970–2023, our metric identifies basic scientists and reveals that their prevalence is higher than expected in prestigious journals, and their association with high impact. They contribute more to conceptualization and writing and tend to create novel combinations of knowledge. We also find their unidirectional career trajectory from basic to applied research, and a systemic applied shift, yet they remain disproportionately central to the world's most influential discoveries. These findings highlight  the critical need to sustain their distinctive role as the intellectual foundation upon which all applied advances are built.

\section{Results}\label{sec:research_design}

\subsection{Application score and its validation }\label{sec:research_design}

\begin{figure}[!t]
  \centering
  \includegraphics[width=\textwidth]{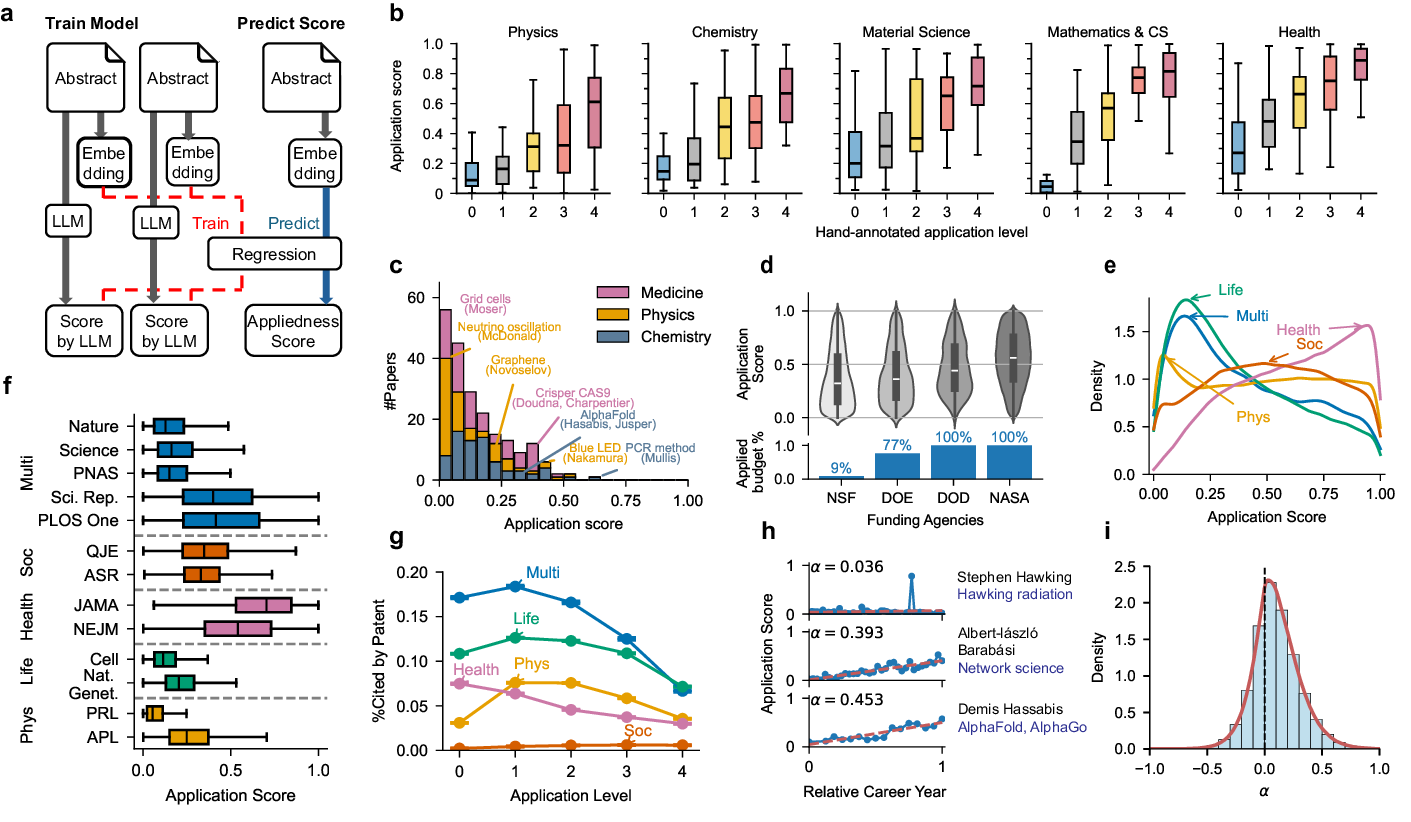}
  \caption{\textbf{Construction and validation of the application score.}
\textbf{a}, Workflow for generating the score. A regression model, trained on GPT-annotated levels and text embeddings from 45k papers, was applied to the embeddings of 62M abstracts. The output is a continuous score normalized to a uniform 0–1 distribution.
\textbf{b}, Predicted scores (y-axis) versus expert-annotated levels (x-axis) across major scientific fields.
\textbf{c}, Histogram of scores for Nobel Prize-winning papers in Medicine (yellow), Physics (blue), and Chemistry (pink).
\textbf{d}, Violin plots of score distributions for papers funded by the NSF, DOE, DOD, and NASA, with corresponding applied research budget shares from \cite{harris2023federal}.
\textbf{e}, Density distributions of the application score by scientific domain (ASJC1).
\textbf{f}, Application score distributions for journals in multidisciplinary, social, medical, life, and physical sciences.
\textbf{g}, The proportion of papers cited by patents (y-axis) against their annotated application level (x-axis), stratified by domain.
\textbf{h}, Application score trajectories over a representative scientist's relative career year (x-axis). A linear model is used to fit the trajectories and trajectories and slopes $\alpha$ are plotted.
\textbf{i}, Distribution of career slopes with KDE fitted curve for slopes $\alpha$ across scientists.}\label{fig1}
\end{figure}

To quantify research orientation, we developed an application score—a continuous, semantically-grounded measure that positions publications along a basic-to-applied spectrum. We derived this score by first prompting an LLM to evaluate a training sample of roughly 45,000 abstracts on a 0–1 scale of appliedness (0 = most basic, 1 = most applied). We then used a regression model to predict appliedness from the text embeddings of these abstracts. This model was then used to assign application scores to the full corpus of 62 million Scopus-indexed publications (1970–2023) (Fig. \ref{fig1}a; see Methods). This approach uses an LLM to identify conceptual distinctions beyond what citation-based proxies can offer, allowing for a more nuanced classification of research orientation across different fields. We tested the method's robustness by varying the prompts, models, and definitions of appliedness, finding high consistency (Spearman $\rho = 0.92-0.98$ ) across ten independent setups. For the final analysis, we selected the configuration that best matched expert assessments (Extended Data Fig.\ref{figS1}).

The score is validated at the paper level. The scores align with expert ratings across five scientific domains with clear basic–applied distinctions (Fig.\ref{fig1}b). Nobel Prize-winning papers—recognized as basic research exemplars—consistently receive low scores, except for some semiconductor and medical applications (Fig.\ref{fig1}c). For example, the 1982 paper on two-dimensional magnetotransport in the extreme quantum limit by D. C. Tsui, H. L. Stormer, and A. C. Gossard~\cite{tsui1982two} received the lowest application score of 0.001346, while the 1985 paper (1993 Chemistry award) on the PCR method by Kary B. Mullis~\cite{saikiEnzymatic1985} ranked among the highest with a score of 0.640. Other well-known breakthroughs that were not basic research include CT imaging (0.388) and the blue LED (0.427).

Aggregated patterns provide further support for validity. The score distributions of papers funded by major U.S. science agencies, for example, align with their publicly stated funding priorities for basic versus applied research~\cite{crs_federal_funding_fy2022} (Fig. \ref{fig1}d). The distribution of application scores across domains (ASJC1; Scopus's 5 major subject domains) and fields (ASJC2; 27 subject fields) (Fig. \ref{fig1}e) reveals expected patterns, with physical and life sciences scoring lower while health and social sciences tend toward applied research. Similarly, the patterns found across the 27 fields align with established expectations (Fig. S2), while confirming that basic research persists in every field \cite{anderson1972more}. The journal-level results in Fig. \ref{fig1}f are consistent with how journals are generally perceived and compared (e.g., Cell versus JAMA).

\FloatBarrier

The relationship between basic and applied research and patent citations differs markedly across disciplines. Analysis of patent citation ratios using SciSciNet data reveals distinct patterns: in the medical sciences, patent citations decrease linearly with the degree of appliedness (Fig. \ref{fig1}g), a finding consistent with prior studies that show basic papers in biomedical domains attract more patent citations than applied ones~\cite{mcmillanAnalysis2000, keTechnological2020}. In contrast, most other fields (with the exception of social sciences) display an inverted U-shaped curve, where patent citations peak for moderately basic research (Fig. \ref{fig1}g). This non-linear relationship, first noted in biotechnology~\cite{wangTales2022}, appears to be a robust finding across domains. This suggests that while medical research follows a direct basic-to-patent pathway, technological application in most other fields benefits most from research of an intermediate basicness.

\subsection{Asymmetric Basic-to-Applied Topic Shift Along Researchers’ Careers}

We also observe that scientific careers frequently undergo transitions from basic to applied research orientations.  Fig.\ref{fig1}h shows representative trajectories of the yearly mean application score. Stephen Hawking ($\alpha=0.036$) remained almost exclusively committed to basic research, with minor applied instances due to credited essays, whereas Albert-László Barabási ($\alpha=0.393$) shifted from statistical physics to network science, and Demis Hassabis ($\alpha=0.453$) from cognitive neuroscience to founding DeepMind, leading applied-oriented breakthroughs such as AlphaGo and AlphaFold. Population-level analysis reveals positive skewness in yearly application score slopes (Fig.\ref{fig1}i), indicating a systemic drift toward application. This aligns with survey evidence that purely basic scientists are a small minority~\cite{bentley2015relationship, sauermannScience2012}, and with longitudinal studies showing declining interest in basic research and rising interest in commercialization~\cite{roachSauermann2017}.

This asymmetry illustrates how scientific careers are structurally open toward applied domains. This may be influenced by evaluation and funding systems that increasingly emphasize translational potential and societal impact~\citep{fang2010lost,rubio2010defining}. Scientists who begin their careers in the basic domain may constitute a specific cohort of scientists, carrying distinctive methods and conceptual resources that later applied work builds upon\cite{agarwal2013industry}. To test this systematically, we classify scientists by the average application score of their first ten publications, assigning the lowest bin as basic, followed by quasi-basic, intermediate, quasi-applied, and applied groups.

\subsection{Basic Scientists’ Positive Effect on Scientific Impact} \label{subsec:impact}

\begin{figure}[!h]
\centering
\includegraphics[width=\textwidth]{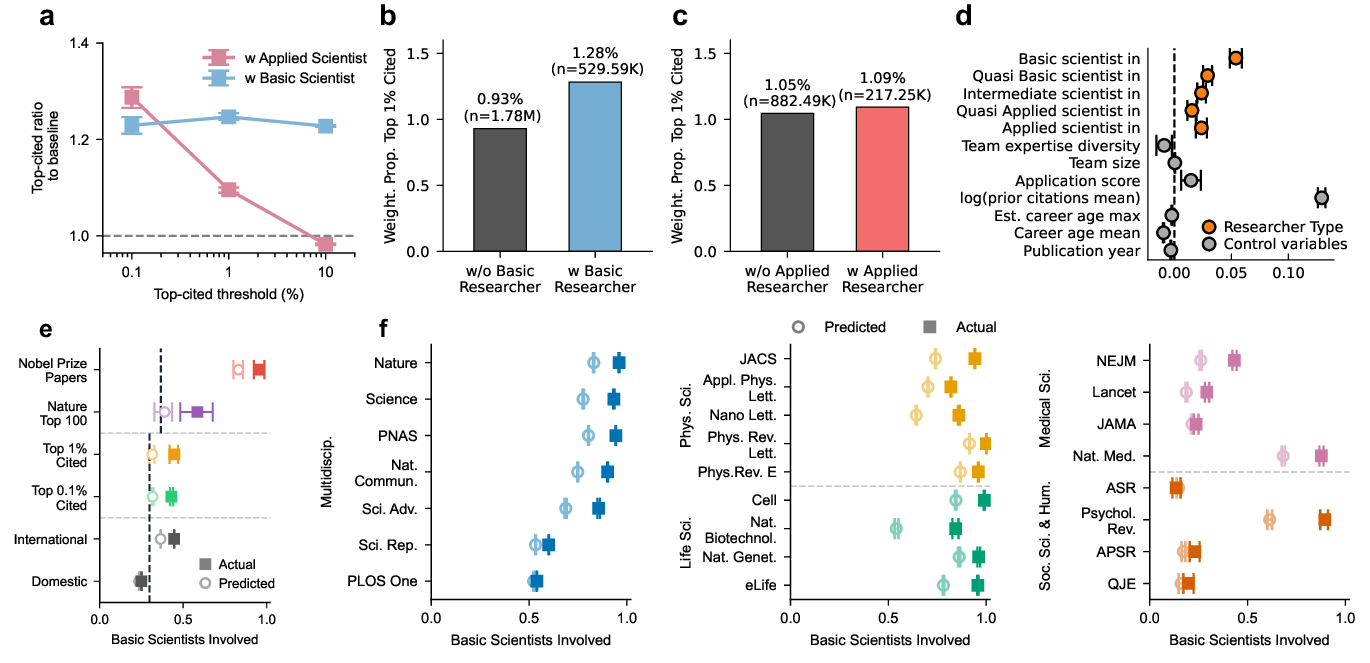}
\caption{\textbf{Basic scientists’ contributions to citation impact.} All analyses use papers published from 1999-2018 and 5-year FWCI.
\textbf{a}, Ratio of top-cited(FWCI) publication rates across citation thresholds (x-axis, percentage of most cited; y-axis, relative ratio), stratified by papers with and without applied or basic scientists. 
\textbf{b}, Weighted proportion of papers in the top 1\% FWCI, adjusted using Coarsened Exact Matching (CEM) (see \ref{subsection:methods_cem}), comparing inclusion and exclusion of basic scientists.
\textbf{c}, Weighted proportion of top 1\% FWCI papers under the same adjustment for applied scientists' involvement.
\textbf{d}, Regression coefficients from models predicting FWCI, with variables for scientists' type and multiple bibliometric covariates (see Appendix C for the regression result).
\textbf{e}, Observed (squares) and predicted (light-colored circles) basic scientist engagement rates across outcome categories (y-axis: Nobel Prize papers, \textit{Nature} Top 100, top 1\% FWCI, top 0.1\% cited) and collaboration types (international, domestic). Dotted horizontal lines indicate the average rate of basic scientist involvement across all publication periods (top) and restricted to 1999–2018 (bottom).
\textbf{f},  Observed and predicted basic scientist engagement rates for prestigious journals and some mega journals.
}\label{fig2}
\end{figure}

The involvement of basic scientists enhances scientific impact, while applied scientists demonstrate no comparable effect. Without controls, papers featuring basic scientists show approximately a 25\% greater likelihood of ranking in the top 1\% of Field-Weighted Citation Impact (FWCI), with similarly elevated probabilities at the 0.1\% and 10\% thresholds (Fig. \ref{fig2}a). Applied scientists produce notable improvements only at the 0.1\% threshold. To account for potential confounding factors—including the generally lower citation rates of basic research papers—we employed coarsened exact matching (CEM) to control for team size, past performance, seniority, knowledge diversity, topic orientation (application score), field (ASJC2), and publication year (see Methods \ref{subsection:methods_cem}). The CEM results reveal that basic scientists' involvement increases the probability of achieving top 1\% citation status by 38\% (Fig. \ref{fig2}b), whereas applied scientists show no substantial effect (Fig. \ref{fig2}c). This pattern holds across top 0.1\% and top 10\% FWCI outcomes (Extended Data Fig. \ref{figS4}). Regression analyses incorporating identical controls further underscore this asymmetry, revealing strong positive associations for basic scientists but small effects for applied scientists (Fig. \ref{fig2}d). Collectively, these findings demonstrate that basic scientists reliably generate higher citation impact, while applied scientists' contributions to such outcomes remain marginal.

We also observe that basic scientists' impact enhancement remains substantial even within applied research, particularly among the top 40\% of application-oriented papers (Extended Data Fig.\ref{figS4}). Furthermore, this finding demonstrates robustness, as basic author participation consistently generates high-impact outcomes when examined through CEM-matched datasets (Extended Data Fig\ref{figS4}). Conversely, applied scientists contribute only marginal FWCI improvements in basic science settings, and notably, applied author involvement actually exhibits detrimental effects (Extended Data Fig\ref{figS4}).

\FloatBarrier

We also find that basic scientists are consistently overrepresented in high-impact scientific outputs. As shown in Fig. \ref{fig2}e, they are disproportionately present in Nobel Prize-winning papers (93.8\%) and in Nature’s list of the 100 most-cited articles (57.0\%) of all time \cite{vannoordenThese2025}. Despite comprising only 20\% of the research population, basic scientists are central to landmark discoveries that shape entire fields. We constructed baselines adjusting for topic composition by application score, field (ASJC2), year, and team size distributions (see \textit{Methods}). Even with these adjustments, basic scientists’ involvement remained above baseline, especially in the top 0.1\% and 1\% cited papers (middle of Fig. \ref{fig2}e). We also find that basic scientists are more likely to be included in international collaboration, and their frequency of presence is higher than expected, suggesting greater engagement in cross-border research.

Basic scientists also show a higher presence in top journals. In leading multidisciplinary journals (\textit{Nature}, \textit{Science}, \textit{PNAS}, \textit{Nature Communications}, and \textit{Science Advances}), over 80\% of papers feature at least one basic scientist (Fig. \ref{fig2}f), far above their 20\% population share. This ratio surpasses what would be anticipated by chance alone. The overrepresentation becomes less marked in journals like Scientific Reports and PLOS ONE. These findings prompt questions about whether basic research is inherently viewed as more innovative or significant—a topic warranting additional study.

\subsection{Basic scientists drive broader and higher impact through intellectual leadership}\label{subsection:mechanism}

\begin{figure}[!h]
  \centering
  \includegraphics[width=1\textwidth]{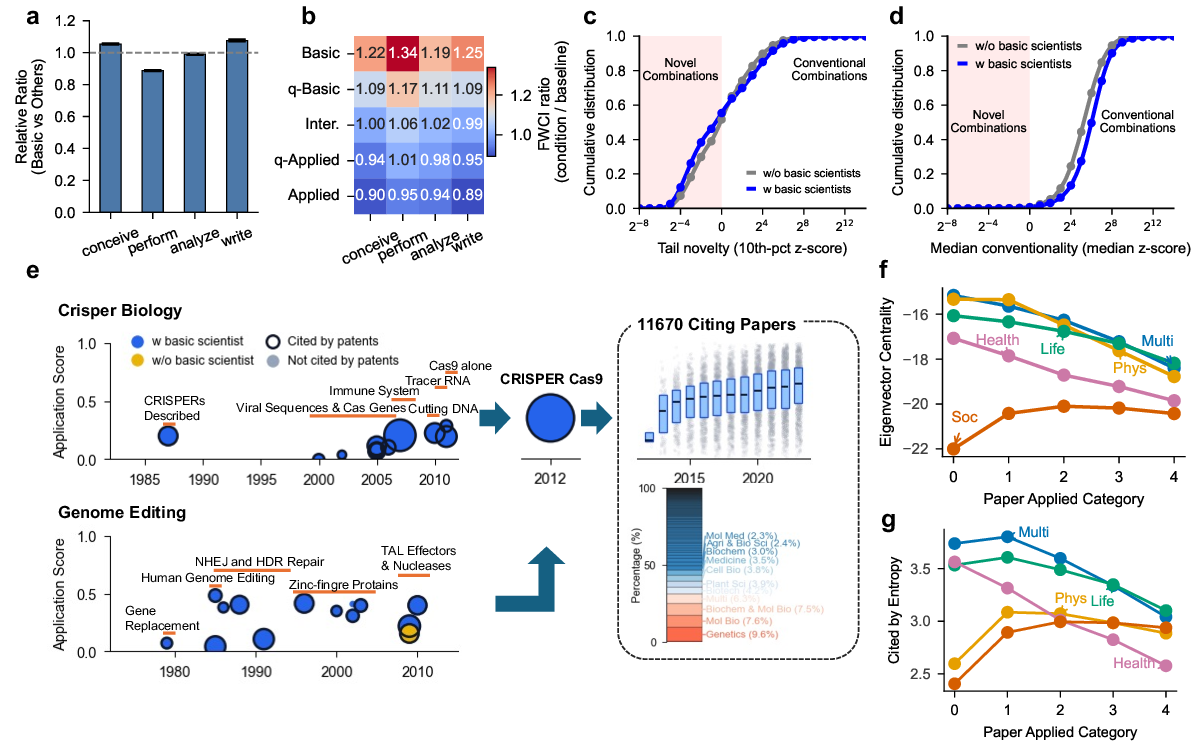}
\caption{\textbf{Basic scientists' engagement in science and pathways of scientific evolution.}
\textbf{a}, Relative task involvement of basic versus non-basic scientists across four roles using author-contribution dataset \cite{xu2022flat}.
\textbf{b}, Associated change in 5-year FWCI for each task by 5-bin career classification of scientists.
\textbf{c,d}, Cumulative distribution of conventionality (median z-score, x-axis) and cumulative distribution of tail novelty of reference paper journal combinations (10th-percentile z-score, x-axis) calculated following \cite{uzziAtypical2013}.
\textbf{e}, CRISPR-Cas9 case study (re-analysis of figures from \cite{doudna2014new}): application scores (y-axis) of landmark publications over time (x-axis) showing CRISPR biology and genome-editing milestones. CRISPR-Cas9 papers are represented with a size indicating citation count. The right panel shows application score distribution for 11,670 citing papers by year and their field (ASJC2) composition.
\textbf{f}, Mean log(Eigenvector centrality) of papers by appliedness category across 5 domains (ASJC1).
\textbf{g}, Mean entropy of citing documents' subfields (ASJC4) by appliedness category across 5 domains (ASJC1).}
  \label{fig4}
\end{figure}

Fig. \ref{fig4}a shows that basic scientists take greater responsibility for writing and conceiving compared to other authors, using structured author-contribution data from Nature, Science, and PNAS (2003–2020)\cite{xu2022flat}. Basic scientists' involvement in any role increases impact, but their performance roles contribute most strongly (Fig.\ref{fig4}b), suggesting that deep engagement drives positive outcomes. Comparing papers with and without basic scientists reveals that teams including them produce more unconventional combinations and novel reference patterns: the cumulative distribution of conventionality shifts leftward, and tail novelty shifts rightward (Fig.~\ref{fig4}c,d). These findings show that basic scientists improve research quality by shaping intellectual design and expanding knowledge recombination.

\FloatBarrier

The CRISPR case demonstrates how scientists involvement shapes the trajectory of basic research (Fig. \ref{fig4}e, which traces the development of CRISPR through a review article \cite{doudna2014new}). Basic scientists led the foundational work on CRISPR technology and genome editing. The high rate of patent citations for this work reveals that these basic research contributions extend beyond the basic domain. While papers in these early chains showed low application scores, the 2012 CRISPR-Cas9 breakthrough\cite{jinekProgrammable2012}—achieved by key basic scientists—sparked a marked increase in appliedness among the 11,670 citing papers (top right of Fig. \ref{fig4}e). This breakthrough has been cited across multiple domains, reaching beyond Genetics and Molecular Biology into Medicine and Plant Science (bottom right of Fig.~\ref{fig4}e).

Consistent with this example, basic research papers—typically authored by basic scientists—show higher eigenvector centrality in the citation network (Fig.\ref{fig4}f), except in the social sciences. In life and medical sciences, these papers draw citations from a wider range of disciplines (Fig.\ref{fig4}g), highlighting their foundational role for diverse applications. This pattern does not appear in the physical and social sciences, likely due to highly basic-oriented papers that form distinct intellectual clusters, such as pure mathematics and fundamental astronomy. These findings indicate that basic scientists not only steer the intellectual direction of science but also establish the foundation for other fields to build upon.

\subsection{Basic scientists generate greater impact from key authorship positions in large, multidisciplinary teams}\label{subsection:formation}

We next turn to a practical question concerning the team configurations under which basic scientists perform best. To address this, we apply stratified Coarsened Exact Matching (CEM) using 5-year FWCI, and top 1\% citation rates independently as outcome variables. Holding other covariates constant, we examine how average treatment effects on the treated (ATTs) vary as a function of a single key factor—such as team size, career age of authors, knowledge diversity or topic orientation—thereby isolating conditions that modulate basic scientists’ contributions while controlling for potential confounders (see Methods).

\begin{figure}[!h]
  \centering
  \includegraphics[width=\textwidth]{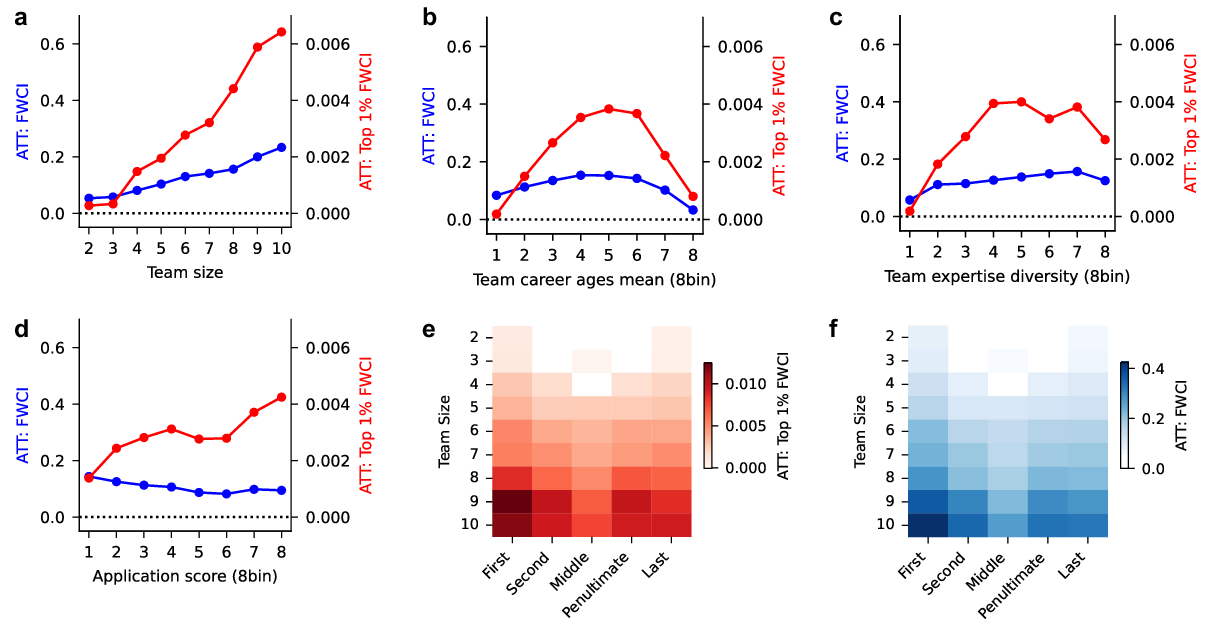}
  \caption{\textbf{Estimated ATT of basic scientists across team contexts.} 
We estimate the average treatment effect on the treated (ATT) of including at least one basic scientist. All analyses use papers published from 1999-2018.
\textbf{a}, ATT estimated across different values of team size. 
\textbf{b}, ATT estimated across bins of team career ages mean. 
\textbf{c}, ATT estimated across bins of team expertise diversity (8 bins). 
\textbf{d}, ATT estimated across bins of application score (8 bins). 
\textbf{e}, ATT on FWCI estimated by author order (first, second, middle, penultimate, last) within each team-size bin. 
\textbf{f}, ATT on top 1\% FWCI rate estimated by author order within each team-size bin. 
Blue lines plot ATT for FWCI (5-year) and red lines plot ATT for the top 1\% FWCI rate; horizontal dotted lines indicate the zero baseline.} 

  \label{fig5}
\end{figure}

The analysis reveals that the involvement of basic scientists has a positive effect on research impact (FWCI, top 1\% FWCI) across all conditions examined. Interestingly, the magnitude of this effect depends on the team context. 
The most pronounced tendency is observed for team size (Fig. \ref{fig5}a): larger teams experience a greater positive impact from the inclusion of basic scientists. 
For seniority of authors (team career ages mean)(Fig. \ref{fig5}b), an intriguing pattern emerges: when the average career age of team members is either too high or too low, the effect of including a basic scientist diminishes, while it reaches its maximum at an intermediate level. 
Regarding knowledge diversity (team expertise diversity)(Fig. \ref{fig5}c), the effect generally increases with higher diversity, although the trend is less pronounced compared to other indicators. 
For topic orientation (application score)(Fig. \ref{fig5}d), the pattern is more complex: as research becomes more application-oriented, the effect on the proportion of top 1\% FWCI papers increases, whereas the effect on FWCI decreases. 
Extended Data Fig.\ref{figS3} presents the effects across fields. 
The inclusion of basic scientists has large positive effects in fields such as Multidisciplinary, Biochemistry, Energy, whereas in fields such as Business, Economics, and Social Sciences, negative effects appear. 

We further classify author positions into five categories and find that the impact is particularly significant when basic scientists occupy key authorship roles—such as the first author—especially in large teams (Fig. \ref{fig5}e, f). These patterns, together with our earlier findings on writing and conceptualization roles (Fig.\ref{fig4}b), suggest that basic scientists achieve maximal influence when situated in leading or mentoring roles within large teams.

\FloatBarrier

\subsection{Applied Drift and the Competition for Basic Scientists}\label{subsec:challanges}

\begin{figure}[!ht]
  \centering
  \includegraphics[width=\textwidth]{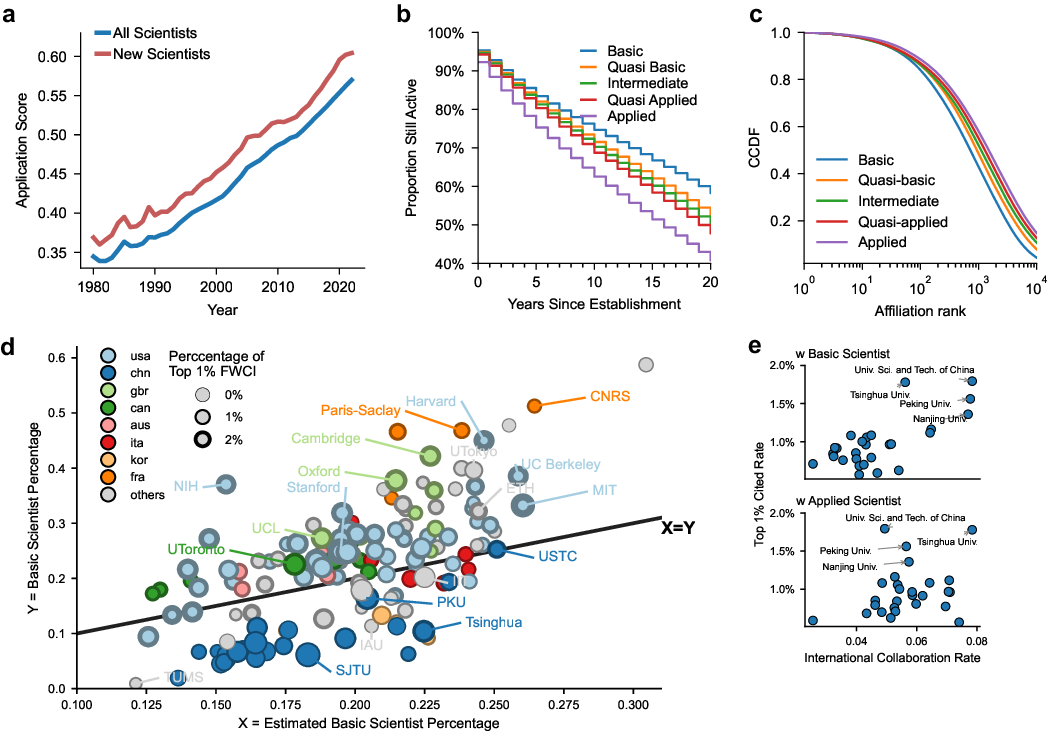}
\caption{\textbf{Career trajectories and institutional distribution of basic scientists.}
\textbf{a}, Application score by year for all scientists with more than 10 papers (blue) and new researchers with more than 10 papers (red).
\textbf{b}, Proportion of scientists still active versus years since establishment (the year of the 10th paper), stratified by classification (Basic, Quasi Basic, Intermediate, Quasi Applied, Applied).
\textbf{c}, For five scientist classification groups(basic to applied), complementary cumulative distribution (CCDF) of number of scientists by affiliation rank (log scale).
\textbf{d}, For top 300 paper-publishing institutions, observed share of basic scientists (y-axis) versus estimated share (x-axis) (see Methods); edge width denotes the percentage of top 1\% FWCI papers and colour indicates country.
\textbf{e}, Among leading Chinese universities, the top 1\% citation rate versus international collaboration rate is shown separately for papers with basic-oriented researchers (top) and applied-oriented researchers (bottom).}
  \label{fig6}
\end{figure}

Basic scientists consistently enhance scientific output quality, yet research and newly joining researchers have shifted toward applied orientations. As illustrated in Fig. \ref{fig6}a (blue line), application scores across all papers have risen over time. Furthermore, scientists who develop sustained publication records (exceeding ten papers) demonstrate greater application orientation from the outset compared to their peers (Fig.\ref{fig6}a, red line), potentially reflecting funding constraints, methodological challenges, or the educational investments required for basic research. Nevertheless, basic scientists continue their work (exceeding 10 papers): survival analysis reveals they remain clustered within specific institutions and exhibit higher persistence rates than other researcher categories (Fig. \ref{fig6}b). These patterns suggest an emerging shortage of basic scientists, despite their essential contributions.

At the institutional level, basic scientists cluster disproportionately within select leading universities (Fig. \ref{fig6}c), particularly at established institutions across the United States, Europe, and Japan (including Harvard, Stanford, and Oxford), as depicted in Fig. \ref{fig6}d. We calculated expected basic scientist proportions for each university by aggregating author contributions according to ASJC categories, application scores, and publication years, then adjusting for institutional publication profiles. After controlling for field composition, many of these high-profile universities maintain substantial basic scientist populations (Fig. \ref{fig6}d).

Chinese leading universities generally host smaller proportions of basic scientists compared to estimates. However, institutions like Peking University, Tsinghua University, Fudan University, and the University of Science and Technology of China achieve internationally competitive outcomes. The proportion of top-1\% FWCI papers shows robust correlation with international collaboration with basic scientist ($r = 0.71$; top panel of Fig.\ref{fig6}e), but only weakly with international applied scientist collaboration ($r = 0.291$; bottom panel of Fig.\ref{fig6}e), indicating that international partnerships with basic scientists critically influence institutional impact.

Globally, the world's most research-intensive universities and institutes—including MIT, Stanford, Harvard, UC Berkeley, Oxford, Cambridge, NIH, and CNRS—combine high concentrations of basic scientists with disproportionate representation among the top-1\% of FWCI papers. NIH, in particular, illustrates how long-term intramural efforts in basic immunology provided the foundation for rapid translation into applied outcomes, most notably the development of COVID-19 vaccines~\cite{corbett2020sars}. These institutions not only sustain pipelines of fundamental research but also act as hubs of international collaboration, reinforcing their central role in shaping frontier science.

\section{Discussion}\label{section:discussion}
Basic and applied scientists shape scientific impact asymmetrically. Prior work emphasized outputs—Bush's framing of basic science as a long-term public investment~\cite{bushEndless1945} or bibliometric analyses showing patents rely disproportionately on high-quality science~\cite{fleming2004science,poege2019science}—but overlooked the scientists themselves. Classifying scientists along a basic–applied spectrum, we find that papers involving basic scientists are 38\% more likely to become top-1\% cited works, an effect that doubles in applied domains. In contrast, applied scientists show no citation benefit in basic science. This asymmetry, where basic scientists elevate cross-field impact without a reciprocal effect, is evident in major breakthroughs: basic microbiology enabled mRNA vaccines, solid-state physics underpinned semiconductors, and basic biology drove CRISPR. Recent analyses confirm that patented inventions link strongly to upstream scientific research, underscoring the role of curiosity-driven work in enabling transformative technologies~\cite{ahmadpoor2017dual}.

Author contributions reveal the mechanisms behind this effect. Basic scientists contribute disproportionately to conceptualization and writing, with the citation impact being strongest when they also perform experimental work (Fig.\ref{fig4}a,b). Their teams create more unconventional combinations of prior knowledge and novel reference patterns (Fig.\ref{fig4}c,d), which aligns with evidence that atypical recombinations of ideas are more likely to produce highly cited work~\cite{uzziAtypical2013}. Basic scientists appear to enhance projects by reframing problems in generalizable terms and connecting disparate literatures—a process supported by cognitive science, which shows that abstract reasoning facilitates cross-domain transfer~\cite{goldwater2016relational}. These effects may stem from basic research training, warranting analysis of links between career formation, education, and performance.

These effects vary by team context and field, strengthening in large, interdisciplinary teams. The impact of basic scientists, shown through stratified matching, grows with team size and disciplinary diversity (Fig.~\ref{fig5}a,c) and is most pronounced when they are the first author (Fig.~\ref{fig5}e,f). This aligns with findings that diverse teams, while often original, can lean toward incremental work~\cite{zhengExpertise2022}. Leadership by a basic scientist may counteract this tendency by guiding projects toward more conceptually bold outcomes. Pasteur’s translational microbiology, revealed through archival analysis~\cite{geison1996private}, and Jumper’s foundational AlphaFold work~\cite{jumper2021highly} exemplify how embedding basic expertise within applied collaborations amplifies novelty and impact.

A career-level asymmetry also emerges: while scientists drift toward application over time (Fig.\ref{fig1}h,i) and new entrants are systematically more applied (Fig.\ref{fig6}a), established basic scientists show the highest survival rates (Fig.\ref{fig6}b). This bifurcation—persistent basic scientists but fewer new entrants—reflects survey evidence that most science PhDs identify with applied work~\cite{sauermannScience2012} and structural pressures from university–industry–government relations~\cite{etzkowitz2000dynamics}. The "valley of death" in translational research creates barriers for returning to basic science~\cite{butler2008translational}, while evaluation systems prioritizing short-term impact exacerbate these dynamics~\cite{rafols2016dominance}. These trends reveal that the shortage of basic scientists stems from declining entry rather than career attrition, underscoring the need for early-career incentives to attract new generations to discovery-oriented science.

Universities have historically protected curiosity-driven research, attracting and retaining basic scientists who often become interdisciplinary leaders~\cite{bushEndless1945}. Funding agencies regard basic research as a public good~\cite{stiglitz1999knowledge} and increasingly promote "transformative research" that reshapes paradigms~\cite{national2019fostering}. Today, basic scientists remain concentrated in prestigious universities, sustaining publishing careers (Fig.\ref{fig6}b) and clustering in top-ranked organizations in the US, Europe, and Japan (Fig.\ref{fig6}c,d), while Chinese universities host fewer, though those with strong international collaborations remain competitive (Fig.\ref{fig6}e). These patterns reflect cumulative advantage, from the Matthew effect~\cite{merton1968matthew,bol2018matthew} to the dominance of elite universities in knowledge production~\cite{altbach2011road}. Across national contexts, a common thread is the institutional commitment to sustaining basic research—through NIH intramural programs in the US~\cite{sampat2012mission}, China’s Thousand Talents program, and the European Research Council (ERC) prioritizing frontier, investigator-driven science~\cite{luukkonen2014european}.

Our study has limitations. The application score, while validated (Fig. \ref{fig1}b–f), reduces complex scientific careers to a single scalar value. Future work should develop multidimensional representations—perhaps vector embeddings—to capture how scientists navigate between basic and applied research. More fundamentally, we must move beyond measuring impact through citations and patents to understanding why basic scientists matter. What mechanisms drive their influence—deeper theoretical insights, longer time horizons, or unique collaboration patterns? How do they contribute through mentorship, methodology development, and "hidden citations" that implicitly shape fields~\cite{meng2024hidden}? Exploring these questions will reveal why institutions continue investing in curiosity-driven research despite uncertain returns, and help us understand the value basic scientists bring to science and society~\cite{bornmann2013societal}.

\section{Methods}\label{section:methods}

\subsection{Research design summary}\label{subsection:methods_summary}
We combine large-scale bibliometric analysis with machine learning to assess contributions of basic scientists. We use the Scopus dataset from 1970-2023 due to its reliable author disambiguation \cite{baas2020scopus}. To compare papers involving basic scientists with others, we restrict analysis to established authors with more than 10 papers. We analyze 62,232,171 papers containing abstracts and classified as Journal Articles, Letters, Reviews, or Conference Papers.

We first develop an application score—a continuous metric positioning publications along the basic–applied research spectrum—through LLM-assisted text embeddings and linear regression (Section~\ref{subsection:methods_score_calculation}), validated using expert ratings and aggregated journal, field, and funding patterns (Section~\ref{subsection:methods_validation}). Using early-career publication profiles, we then identify basic scientists (Section~\ref{subsection:methods_identifying}). With this classification, we evaluate their influence on scientific impact via FWCI and top-cited publication rates (Section~\ref{subsection:methods_impact}). Analyses span the full corpus and application-oriented subsets, controlling for team size, past performance, seniority, knowledge diversity, topic orientation (application score), field (ASJC2) and publication year. We also examine basic scientists' participation in high-impact science by calculating baseline probability of their involvement (Section~\ref{subsection:methods_highimpactscience}).

To identify underlying mechanisms, we analyze structured contribution statements (Section~\ref{subsection:methods_roles}), citation network dynamics, and team roles. We assess how team size, seniority (mean career age of authors), knowledge diversity, topical orientation (application score) affect basic scientists' influence, yielding insights for optimal team composition (Section~\ref{subsection:methods_formation}).

\subsection{Calculating application score}\label{subsection:methods_score_calculation}
To quantify the position of scientific publications along the basic–applied research spectrum, we construct a continuous application score using a machine learning approach guided by large language model (LLM) annotations. The process comprises three steps: text embedding of abstracts, LLM-based score annotation, and corpus-wide prediction.

We begin with a corpus of 62,232,171 Scopus-indexed publications with available abstracts published between 1970 and 2023. Each abstract is embedded into a 1,536-dimensional vector space using OpenAI’s \texttt{text-embedding-3-small} model, which captures the semantic content of the text in a dense numerical representation.

A random subset of 45,000 publications is then annotated by an LLM to assign a training score from 0 (most basic) to 1 (most applied), in 0.1 increments, using a prompt designed to evaluate the appliedness of each abstract. The resulting LLM-generated scores are paired with their corresponding embeddings to train a linear regression model, which is then applied to predict application scores for the full publication corpus. Predicted scores are normalized to a uniform distribution over the [0, 1] interval to ensure boundedness and interpretability.

We evaluated two LLMs (GPT-4o and Gemini 2.0 Flash) across five prompts, and selected the optimal combination based on alignment with expert annotations. The final configuration was GPT-4o with Prompt 5 (other prompts are listed in Appendix B), shown below.

\begin{mdframed}[linewidth=1pt, linecolor=gray, backgroundcolor=white]
\textbf{Prompt 5 (From \textit{The New ABCs of Research}\cite{shneiderman2016new}) }
Here are the characteristics of applied research: 

1. Applied research is mission driven:  
Applied research is based on scientists working closely with practitioners to understand natural and made-world problems. Taking on real problems requires taking the time to learn the practitioners’ language and forces applied scientists to address problems as defined by others. Applied research is sometimes described as being mission driven, a description that suggests that it always has a practical goal, unlike curiosity-driven basic research. Applied scientists are more often attracted to solving problems that have social, health, or economic impacts. 

2. Applied research looks for practical solutions and guidelines:  
Promoters of applied research are often satisfied with producing workable solutions and practical guidelines, but are even happier if the solution can achieve broad applicability. They are especially thrilled to see the prompt application of their work to goals such as increasing agricultural yields or raising manufacturing quality. 

3. Applied research examines complex interactions between multiple variables:  
Applied problems are often complex, with many variables that cannot be easily controlled because of rich and changing context, and are sometimes called “wicked problems,” that is, problems for which solutions sometimes work and sometimes fail, because of changing conditions. Applied scientists strive to understand interactions between variables such as a mayoral candidate’s political positions, media expenditures for positive versus negative advertising, and socioeconomic variables as they influence voter turnout across the hundreds of polling places in a large city. Applied scientists like conquering complexity and learning how to deal with changing conditions. Sometimes solutions in complex situations may not come with causal explanations, and therefore these solutions need revision when conditions change. 

4. Applied research uses realistic (rather than idealized) scenarios:  
Applied scientists enjoy working on realistic situations, trying multiple solutions, refining promising ideas, and getting their hands dirty. Applied scientists thrive on these challenging situations, which force them to learn more, try many solutions, and fail frequently until they find success. Simplified problems (“toy problems”) may be too easy for them, not offering enough of a challenge or appearing to be merely games.

Here are the characteristics of basic research: 

1. Basic research is curiosity driven:  
Basic research is generally understood to be motivated by curiosity or a drive to understand the world we live in, rather than the need to solve an existing problem. It stems from observations of the world, a desire to organize knowledge, and an eagerness to predict how the world will behave. For example, questions about the natural world might include, are there animals with an odd number of legs? What chemicals efficiently store energy? What gases are on Jupiter? Questions about the made world might include, what is the impact of financial incentives on getting individuals to change their exercise habits? How can domestic conflicts be resolved before they become violent? Is there necessarily a trade-off between healthcare quality and costs? While most observers are happy to let individuals follow their curiosity, when they or their organizations seek public funds, valid questions include, how much curiosity-driven exploration is appropriate? A central criterion is whether answers are actionable; will research outcomes lead to changes that bring nearterm economic or other benefits? 

2. Basic research employs reductionist models:  
A second common feature of basic research is the reductionist model, which presumes that phenomena can be studied by changing one variable at a time. Basic scientists may be naturalists exploring forests, but basic research is often tied to laboratory situations where conditions can be controlled to limit variability and promote replicability. Basic scientists like the process of solving one problem at a time, in the belief that putting together independent results can explain larger interconnected phenomena. For example, a chemist can study the conductivity of electricity in a metal while controlling the room temperature, altitude, and magnetic field. The results from other chemists can then be combined to make formulas that have the fewest variables needed to predict what might happen in novel situations. 

3. Basic research searches for universal principles:  
Proponents of basic research often claim that it leads to general theories and predictions. Tracking the movements of Mars was meant to lead to a general theory of planetary motion, and studying Escherichia coli is meant to produce knowledge that is true for many bacteria. Promoters of basic research believe that their work will result in broad understanding of multiple phenomena. Critics argue that a narrow focus on special cases chosen in controlled experiments may not generalize when the results are applied more widely. 

4. Basic research relies on simplifications and idealizations:  
A fourth feature of basic research is that the use of simplifications or abstractions of complex phenomena is acceptable in order to facilitate research. Physicists study gravitational attraction between bodies by assuming that the mass of each body is concentrated in a single point; similarly, social network theorists assume all nodes in a graph have the same properties. These simplifications and the use of synthetic data with uniform or normal distributions, rather than real data with unusual distributions, make research easier and support application to other idealized problems. Basic scientists see these simplifications as clean, clear problems that yield elegant solutions. 

Now, you will be provided with abstracts of academic publications. Based on the above characteristics, you are to rate each publication based on whether they are basic or applied research on a continuous scale from 0 to 1. A score of 0 means the research is completely basic. A score of 1 means the research is completely applied. A score of 0.5 means the research has equal basic and applied elements. 
\end{mdframed}

\subsection{Validating score accuracy}\label{subsection:methods_validation}

\methodhead{Human labelling} 
To validate our scores against expert judgment, we selected four domains with clear basic–applied distinctions: Physics, Chemistry, Mathematics \& Computer Science, and Medicine \& Pharmaceutics. From each domain, we randomly sampled 200 publications using Scopus All Science Journal Classification (ASJC) codes. A domain expert for each field then independently rated each paper’s application level on a one-to-five scale (1 = most basic; 5 = most applied). We then compared our model's application scores against these human ratings.

\methodhead{Nobel Prize-winning papers}
We started with the list of Nobel Prize–related papers through 2016 compiled by \cite{li2019dataset}, which we matched to our Scopus database (1970-2023). To this list, we added papers from the 2017-2023 prizes (38 papers), yielding a total of 212 papers (71 in Chemistry, 67 in Physics, and 74 in Medicine).

\methodhead{Agency funding profiles}
We also compared our application scores with the research priorities of four major U.S. federal funding agencies. Using the 2023 congressional report~\cite{harris2023federal}, we obtained the basic versus applied budget allocations for the National Science Foundation (NSF), Department of Energy (DOE), Department of Defense (DoD), and NASA. We then identified 236,259 publications from that year with funding acknowledgements from these agencies in the Scopus data.

\methodhead{Patent citation patterns} 
Using patent citation data from SciSciNet~\cite{lin2023sciscinet}, which contains OpenAlex IDs for patent-cited papers, we retrieved the DOI of each paper from the OpenAlex database and linked it to its Scopus ID. This process yielded 2,368,194 papers cited by patents published after 1998.

\subsection{Identifying basic scientists}\label{subsection:methods_identifying}
Our analysis covered 2,490,638 scientists with 10 or more publications in Scopus (1970–2023), hereafter referred to as \textit{established authors}, encompassing their 63,249,508 associated publications. We characterized each scientist's early-career profile by averaging the application scores of their first 10 publications. Scientists were then ranked by this average score and divided into five equal-sized categories: basic (lowest), quasi-basic, intermediate, quasi-applied, and applied (highest). This classification was fixed at the early-career stage to maintain temporal consistency.

\subsection{Evaluating basic scientists' asymmetric impact}\label{subsection:methods_impact}

Publications were grouped by presence or absence of basic scientist authorship. We calculated 5-year Field-Weighted Citation Impact (FWCI) for papers published before 2019 and top-cited rates (top 10\%, 1\%, 0.1\%) for all papers. Comparisons were conducted on: (i) all papers authored by study scientists; (ii) papers with top 40\% application scores (most applied research); and (iii) papers with bottom 40\% application scores (most basic research). For reference, we compared basic scientists' impact against papers with applied scientist authorship.

\subsection{Coarsened Exact Matching}\label{subsection:methods_cem}
In this study, we estimate the effect of including a Basic scientist (or an Applied scientist in Fig. \ref{fig2}c as a coauthor on research impact (FWCI over five years and top 1\% FWCI) using Coarsened Exact Matching (CEM) \cite{iacusCausal2012}. We define the treatment indicator as $D_i \in \{0,1\}$, where $D_i=1$ if the focal scientists type is included and $D_i=0$ otherwise. The outcome variable $Y_i$ is FWCI over five years or top 1\% FWCI (binary).
The covariates used for matching are described below.

\methodhead{Team size (\texttt{team size})} 
Prior studies have shown that larger teams tend to produce more highly cited research, partly due to increased resources, specialization, and visibility~\cite{guimeraTeam2005,larivireTeam2014,wuchtyIncreasing2007}. This raises the possibility that team size may drive the observed citation advantage of papers involving basic scientists.

\methodhead{Past performance (\texttt{team prior citations mean}, \texttt{number of established authors})} 
The Matthew effect in science suggests that more experienced or highly cited scientists tend to benefit from cumulative citation advantage, earning more citations over time~\cite{merton1968matthew}. In this case, the observed citation advantage may reflect the team’s past success rather than basic scientists’ additive contribution to the focal paper. We therefore adjust for the average cumulative citations of team members (measured at the end of the calendar year prior to publication). In CEM analysis, we additionally include \texttt{number of established authors}, defined as the team size with more than 10 publications in the past and has its scientist type identified. \texttt{number of established authors} is not included in regression due to multicolinearity.

\methodhead{Seniority (\texttt{team career ages mean, established author career ages max})}
It has been shown that scientists' career age is associated with their performance~\cite{gingrasEffects2008, liEarly2019,sunaharaUniversal2023,linRemote2023}. Basic scientists' impact, including their leadership, may also be a result from their seniority. As such, we use \texttt{team career ages mean} to control for the age structure of the team, and \texttt{established career ages max} to control for the experience of the established author who is likely to serve as a supervisor within the team.

\methodhead{Knowledge diversity (\texttt{team expertise diversity})} 
Recent work~\cite{zhengExpertise2022} has shown that teams with greater expertise diversity tend to produce higher-impact research. Since the presence of basic scientists may itself contribute to such diversity, part of the observed citation advantage could reflect the broader knowledge base of these teams rather than intrinsic attributes of basic scientists. To examine this possibility, we follow the method proposed in that study and compute \texttt{team expertise diversity} as the average pairwise cosine dissimilarity among coauthors’ expertise vectors, derived from their cumulative publication records. Full methodological details are provided in the original study \cite{zhengExpertise2022}.

\methodhead{Topic orientation (\texttt{application score})} 
Citation potential varies across research topics due to inherent differences in novelty, maturity, or practical significance. It is possible that basic scientists preferentially engage with topics more likely to be cited—or that basic research topics themselves, regardless of authorship, attract more citations. To address this, we include each paper’s \texttt{application score} as a covariate. This allows us to distinguish scientist-level effects from topic-level citation dynamics along the basic–applied spectrum. 

\par\vspace{1.5ex}\noindent
Analyses additionally include \texttt{publication year} and \texttt{field} (ASJC2).

\par\vspace{1.5ex}
In the CEM procedure, all covariates except for \texttt{field} and \texttt{publication year} are coarsened into eight quantile bins (\texttt{field} is excluded in Fig.\ref{fig5} to avoid small stratum sizes). This coarsening induces strata $s$, each consisting of observations with identical coarsened values, and strata containing only treated or only control units are discarded. We denote the number of treated units in stratum $s$ by $m_s^T$ and the number of control units by $m_s^C$, with $m_T=\sum_s m_s^T$ and $m_C=\sum_s m_s^C$. Following equation (6) of \cite{iacusCausal2012}, the weights are defined as
\[
w_i =
\begin{cases}
1, & i \in T_s \ (\text{treated in stratum } s), \\[6pt]
\dfrac{m_C}{m_T}\cdot\dfrac{m_s^T}{m_s^C}, & i \in C_s \ (\text{control in stratum } s), \\[6pt]
0, & \text{otherwise}.
\end{cases}
\]
Based on these weights, the weighted mean outcomes are defined as
\[
y_{\text{treated}} = \frac{\sum_i w_i D_i Y_i}{\sum_i w_i D_i}, \qquad
y_{\text{control}} = \frac{\sum_i w_i (1-D_i) Y_i}{\sum_i w_i (1-D_i)}.
\]
Note that for the binary outcome top 1\% FWCI, $y_{\text{treated}}$ and $y_{\text{control}}$ represent weighted proportions. In Fig. \ref{fig2}b,c, we visualize $y_{\text{treated}}$ and $y_{\text{control}}$.

The average treatment effect on the treated (ATT) is then estimated as
\[
\widehat{\text{ATT}} = y_{\text{treated}} - y_{\text{control}}.
\]
Stratified CEM, introduced in Section~\ref{subsection:methods_formation}, is used to examine optimal team configurations and further support the robustness of the observed citation advantage across a wide range of team contexts (Fig.~\ref{fig5}).

\subsection{Regression analysis}\label{subsection:methods_regression}
\methodhead{Regression analysis}
To examine the effect of the inclusion of basic/applied scientists as co-authors on the scientific impact (measured by citations), we conducted an OLS regression analysis. The dataset covered a 20-year period from 1999 to 2018 to ensure accurate calculation of citation counts. To avoid bias from extremely large collaborations, papers with more than 20 co-authors were excluded, and a 1\% random sample of the remaining papers was used ($n = 297,613$).

As the dependent variable, we employed the field-weighted citation impact (FWCI) based on the number of citations received within five years of publication. To approximate a normal distribution, FWCI was log-transformed (log(FWCI + 1)).

As intervention variables, we constructed dummy variables (\texttt{Basic author in}, \texttt{Quasi Basic author in}, \texttt{Intermediate author in}, \texttt{Quasi Applied author in}, \texttt{Applied author in}) indicating whether the focal paper included co-authors of different established author types (see \ref{subsection:methods_identifying}).

Control variables included the following factors that may affect the dependent variable ($fwci$): \texttt{team size}, \texttt{team expertise diversity}, \texttt{application score}, \texttt{log(team prior citations mean)}, \texttt{team career ages mean}, \texttt{established author career ages max}.
The covariates used in the regression were essentially the same as those used in the CEM analysis; however, to prevent multicollinearity, variables with a VIF of 5 or higher were excluded (specifically, \texttt{field} and \texttt{number of established authors}).
Standard errors were clustered at the first-author level to account for correlation in outcomes among papers written by the same scientists.

\subsection{Evaluating basic scientists' prevalence in high-impact science}\label{subsection:methods_highimpactscience}
We assessed basic scientist engagement rates (proportion of papers with $\geq$1 basic scientist) across high-impact publication sets. These included: (i) landmark papers—Nobel Prize-winning papers, \textit{Nature}'s 100 most-cited papers~\cite{vannoordenThese2025}, and top 0.1\% most-cited papers in Scopus; (ii) internationally coauthored papers (defined as no single country representing majority of affiliations) versus single-country papers; and (iii) prestigious journals, including \textit{Nature} Index journals and leading social science titles.

To account for potential bias from topic and team size distributions, we calculated baseline engagement rates for each paper set $k$:
\[
\textit{baseline}_k = \frac{\sum \left( n_{i,j,k} \cdot \frac{b_{i,j}}{n_{i,j}} \right)}{\sum n_{i,j,k}}
\]
where $n_{i,j,k}$ represents papers in set $k$ with application score in bin $i$ (1--10) and team size $j$ ($\geq$10 authors grouped together); $n_{i,j}$ and $b_{i,j}$ represent total papers and papers with basic scientists, respectively, stratified by the same bins across the entire database.
```

\subsection{Identifying basic scientists' functional roles}\label{subsection:methods_roles}

We examine functional contribution patterns using a structured author contribution dataset\cite{xu2022flat}, which covers papers published in \textit{Nature}, \textit{Science}, and \textit{PNAS}. The dataset records four self-reported task roles: conceiving the research, performing experiments, analyzing data, and writing the manuscript. Author identities are matched to Scopus records to determine each contributor’s basic–applied classification, allowing us to identify role tendencies more frequently associated with basic scientists.



\subsection{Detailed analysis of basic scientist impact using stratified CEM}\label{subsection:methods_formation}

To provide practical guidance on how to deploy basic scientists most effectively, 
we estimate heterogeneous treatment effects under varying team contexts using 
stratified Coarsened Exact Matching (CEM). In this approach, we fix one contextual 
variable—such as \texttt{team size}, \texttt{team career ages mean}, \texttt{application score}, 
or \texttt{team expertise diversity}—and estimate the average treatment effect on the 
treated (ATT) of basic scientist involvement within each stratum. For each context value, 
we compare papers with and without basic scientist involvement using CEM. Matching is 
performed on the remaining covariates to ensure comparability between groups. This allows us to isolate the contextual conditions under 
which basic scientists are most impactful (Fig.~\ref{fig5}a--d).

We further examine how a basic scientist’s position in the author list—used here as a proxy 
for their role—shapes their contribution to scientific impact. Here we use two contextual 
variables: \texttt{team size} and \texttt{basic scientist position}. \texttt{Basic scientist position} 
classifies papers based on what position or positions the basic scientists occupy in the paper:
\textit{first}, \textit{second}, \textit{penultimate}, \textit{last}, or \textit{middle} (which includes all 
other positions). For example, papers with five authors and a basic scientist in the first author 
position are matched against papers with five authors but no basic scientist on the author team. 
These analyses assess whether certain author roles amplify the contribution of basic scientists 
beyond their general presence on the team (Fig.~\ref{fig5}e,f).



\subsection*{Data availability}

The raw Scopus data used in this study are not publicly available due to commercial licensing restrictions. Our annotation datasets will be made available upon reasonable request. All other data sources used in our analyses are openly accessible: Microsoft Academic Graph (MAG) raw data are publicly available at \href{https://docs.microsoft.com/en-us/academic-services/graph/}{https://docs.microsoft.com/en-us/academic-services/graph/}. Structured author contribution data are publicly available at \href{https://zenodo.org/records/6569339}{https://zenodo.org/records/6569339}. Patent citation data are available through the SciSciNet\cite{lin2023sciscinet}.



\bibliography{sn-bibliography}

\backmatter

\bmhead{Supplementary information}
Prompts and paper-level application scores are available in \textit{Supplementary Information}.

\bmhead{Acknowledgments}
We are especially grateful to W. Huang for his extensive input across some stages of the research. We thank M. Gonokami and T. Someya for their insights on practical implications. We also thank C. Miura, M. Kakizoe, S. Lee, R. Sato, Y. Mei, Y. Onishi, and various conference audiences for their helpful comments and discussions.
This work was supported by JSPS KAKENHI Grant Numbers 25K21519 and 21K19817. 

\bmhead{Author contributions}
K.A considered the original idea of the paper. R.K. and K.A. collaboratively conceived the study and reviewed all the results. K.A. conceived and calculated \textit{application score}. R.K., K.N., B.M., and K.A. prepared and analysed the data. K.N. led the regression and causal analysis. B.M. led the analysis on Microsoft Academic Graph (MAG) data and the preparation of some key inputs. I.S. provided insights on policy implications. R.K. managed the project and drafted the manuscript. K.A. and I.S. supervised the project. K.A. and K.N. and revised the manuscript. All authors contributed to the interpretation of data and revisions.

\bmhead{Competing interests}
The authors declare no competing interests.

\bmhead{Additional information}
Correspondence and requests for materials should be addressed to Kimitaka Asatani.

\begin{appendices}

\clearpage
\section*{Extended Data Figures}\label{sec:extended_figures}

\begin{supplementfigure}[ht]
  \centering
  \includegraphics[width=\textwidth]{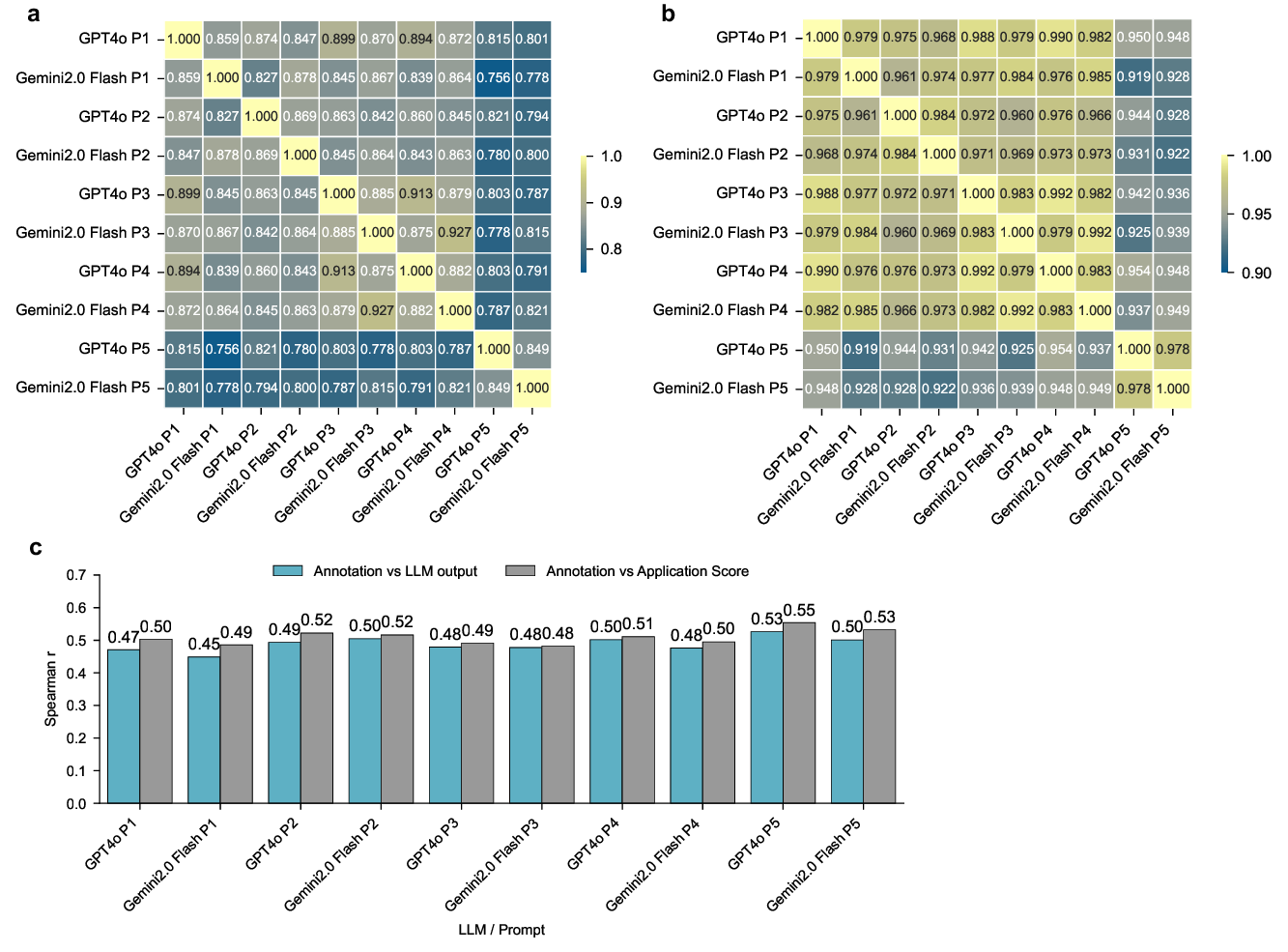}
  \caption{\textbf{Robustness of application scores to prompt design and LLM choice.} 
  We assess score robustness across five prompts and two large language models (GPT-4o and Gemini 2.0 Flash). 
  (\textbf{a}) Spearman correlations of LLM-annotated scores across models and prompts. 
  (\textbf{b}) Spearman correlations of final scores trained on LLM annotations across models and prompts. 
  (\textbf{c}) Spearman correlations of human annotations with LLM scores (two LLMs × five prompts) and final scores. }
  \label{figS1}
\end{supplementfigure}
\clearpage

\begin{supplementfigure}[ht]
  \centering
  \includegraphics[width=\textwidth]{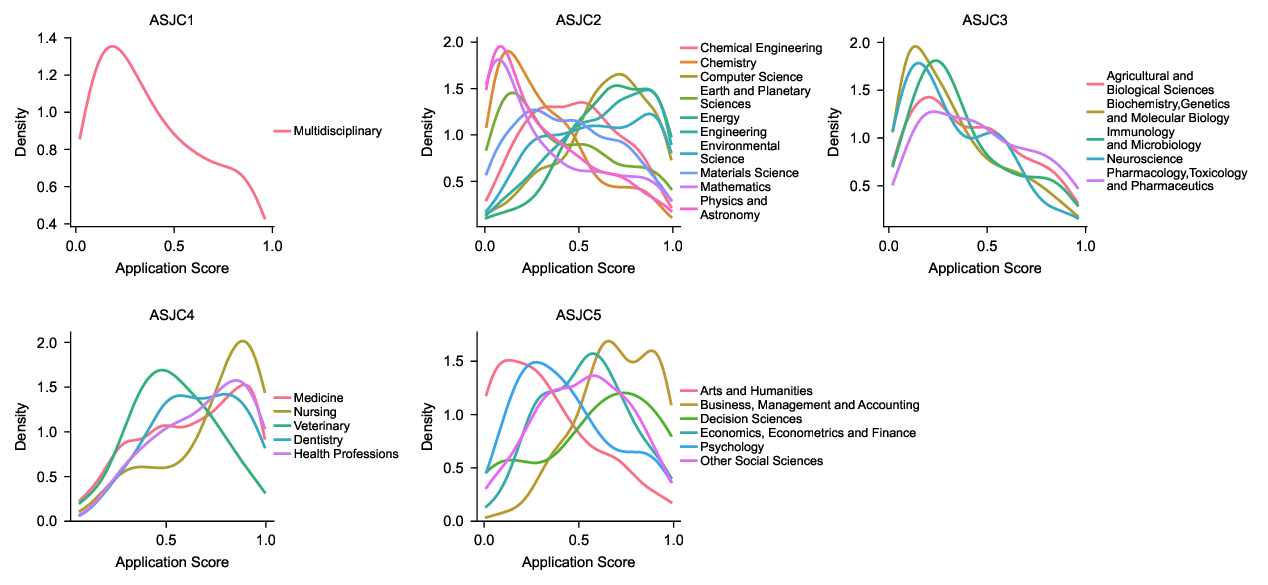}
  \caption{\textbf{Distribution of application scores across fields (ASJC2)} Kernel density estimates of application score distributions are shown for subfields within five major ASJC domains (ASJC1). Each curve represents the distribution for a specific field (ASJC2), with colors distinguishing subfields. The y-axis indicates density, normalized within each panel. Legends to the right of each panel list the corresponding fields (ASJC2).}
  \label{figS2}
\end{supplementfigure}

\begin{supplementfigure}[ht]
  \centering
  \includegraphics[width=\textwidth]{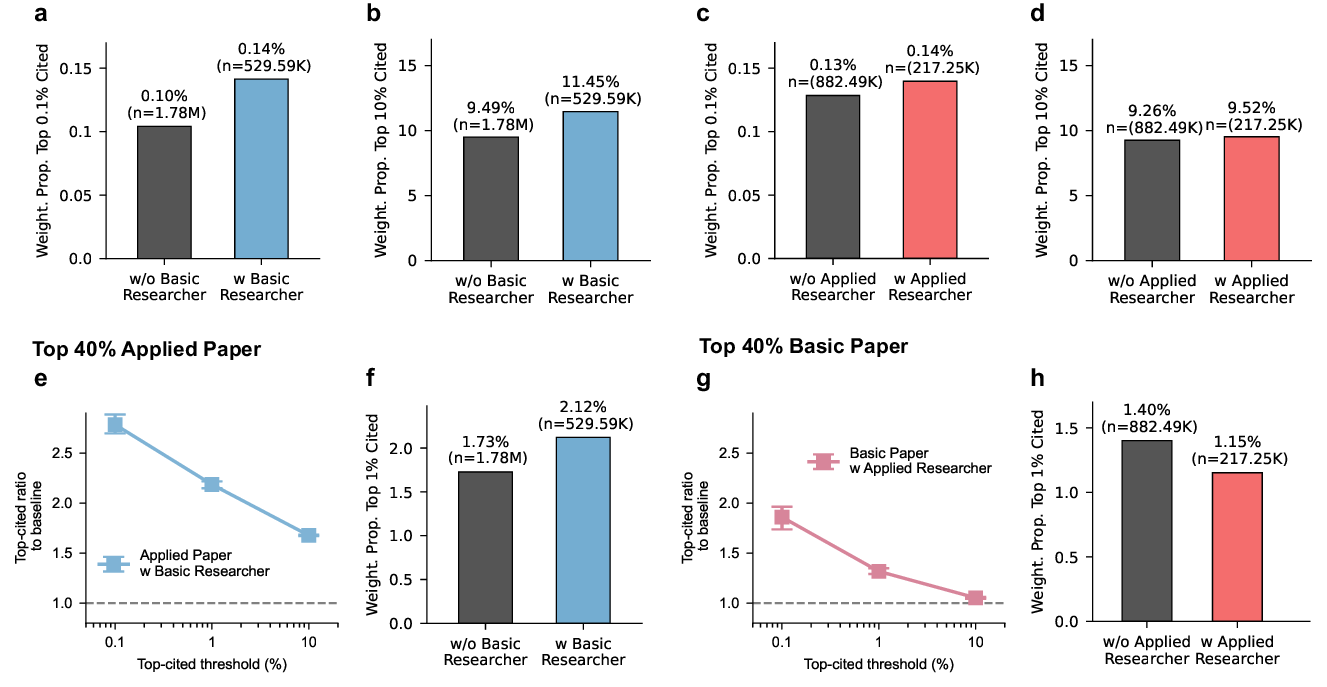}
\caption{\textbf{Detailed analysis of the impact of basic and applied researchers.}  All analyses use papers published from 1999-2018 and 5-year FWCI.
\textbf{a}, Proportion of papers in the top 0.1\% FWCI, adjusted using Coarsened Exact Matching (CEM) (see \ref{subsection:methods_cem}), comparing inclusion and exclusion of basic scientists. 
\textbf{b--d}, Same as in (a), but for the top 10\% FWCI (b), and for inclusion/exclusion of applied scientists in the top 0.1\% (c) and top 10\% (d). 
\textbf{e}, Ratio of top-cited publication rates across citation thresholds (x-axis, percentage of most cited; y-axis, relative ratio to baseline), stratified by applied papers with and without basic scientists. 
\textbf{f}, Proportion of top 1\% FWCI applied papers, adjusted using CEM, comparing inclusion and exclusion of basic scientists. 
\textbf{g--h}, Same as in (e--f), but restricted to basic papers with and without applied scientists. 
Together, these results highlight that basic scientists disproportionately increase the likelihood of high citation impact even in applied domains, whereas applied scientists do not.}
  \label{figS4}
\end{supplementfigure}

\begin{supplementfigure}[ht]
  \centering
  \includegraphics[width=\textwidth]{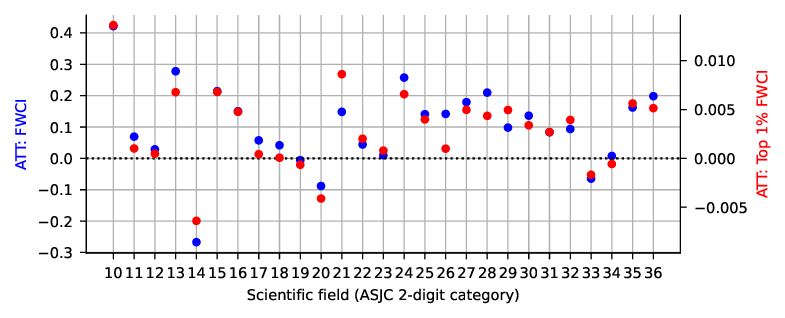}
  \caption{\textbf{Estimated ATT of basic scientists' involvement across fields (ASJC2)..} Papers are stratified by the fields (ASJC2), and within each field we estimate the average treatment effect on the treated (ATT) of basic scientists involvement. For each field, ATT is calculated using two outcome measures: FWCI (blue) and top 1\% FWCI (red). Representative field (ASJC2) include 10 (Multidisciplinary), 13 (Biochemistry), and 21 (Energy), whereas negative effects are observed in fields such as 14 (Business), 20 (Economics), 33 (Social Sciences).
  } 
  \label{figS3}
\end{supplementfigure}

\clearpage
\section*{Appendix}

\subsection*{Appendix A: Validating score robustness against LLM and prompt choice}\label{subsection:methods_llm_robustness}

To assess our method's robustness, we tested combinations of two LLMs (GPT-4o and Gemini 2.0 Flash) with five distinct prompts. The prompts varied; some provided no definition, while others cited definitions from academic and empirical literature~\cite{bushEndless1945,schauzWhat2014,oecd2015frascati,shneiderman2016new}.

We evaluated the consistency of both the raw and final scores from these combinations using Spearman's rank correlation. The LLM-prompt combination used in this study was selected because its final scores showed the highest correlation with a human-annotated training set of 1,000 papers from five scientific domains. As detailed in Extended Data Fig.~\ref{figS1}, the chosen combination was GPT-4o with Prompt 5. All prompts are available in the Supplementary Information.

The results, shown in Extended Data Fig.~\ref{figS1}, reveal three key aspects of the method's robustness. First, the raw outputs from all LLM-prompt combinations are highly correlated ($\rho \approx 0.77$–$0.88$), indicating consistent initial scoring (Panel A). Second, after these outputs are used to train our final scoring models, the correlations between the models' predictions increase significantly ($\rho > 0.90$; panel B), which shows that the training process reduces initial inconsistencies and produces highly stable final scores (Panel B). Finally, when compared against human annotations, both the raw LLM scores and the final model scores show moderate agreement  ($\rho \approx 0.45$–$0.55$; panel C), suggesting that the automated scores serve as a useful approximation for human judgment at scale (Panel C).

Collectively, these findings confirm that our scoring framework is robust to the choice of LLM and prompt, and that the selected GPT-4o and Prompt 5 combination offers both stability and strong alignment with human evaluation.

\subsection*{Appendix B: Prompts of scoring appliedness from paper abstract}\label{A_prompts}

\begin{mdframed}[linewidth=1pt, linecolor=gray, backgroundcolor=white] \textbf{Prompt 1 (No explicit definition)}
You are an excellent scientist familiar with basic and applied research in any field. Please score with continuous values, using 0 for basic research and 1 for applied research, with the following abstract. A score of 0 means fundamental research, a score of 0.5 means research that contains both basic and applied topics, and a score of 1 means fully applied research. \end{mdframed} 

\begin{mdframed}[linewidth=1pt, linecolor=gray, backgroundcolor=white] \textbf{Prompt 2 (Empirical definition of understanding vs. practicality\cite{bushEndless1945})}
You are a great generalist scientist familiar with scientific theories and applications in every field. You will be provided with abstracts of academic publications. You are to rate each research based on their basic-applied orientation on the continuous range between 0 and 1. A score of 0 means the research is completely basic, focusing on fundamental understanding without any potential for practical application. A score of 1 means the research is completely applied, with practical applications immediately in sight. A score of 0.5 means the research has equal basic and applied elements. \end{mdframed} 

\begin{mdframed} \textbf{Prompt 3 (From \textit{Science: The Endless Frontier})\cite{schauzWhat2014}} Basic research is performed without thought of practical ends. It results in general knowledge and an understanding of nature and its laws. This general knowledge provides the means of answering a large number of important practical problems, though it may not give a complete specific answer to any one of them. The function of applied research is to provide such complete answers. You will be provided with abstracts of academic publications. You are to rate each publication based on whether they are basic or applied research on a continuous scale from 0 to 1. A score of 0 means the research is completely basic. A score of 1 means the research is completely applied. A score of 0.5 means the research has equal basic and applied elements. \end{mdframed}

\begin{mdframed}
\textbf{Prompt 4 (From OECD \textit{Frascati Manual}\cite{oecd2015frascati})} Basic research is experimental or theoretical work undertaken primarily to acquire new knowledge of the underlying foundation of phenomena and observable facts without any particular application or use in view. Applied research is original investigation undertaken in order to acquire new knowledge. It is, however, directed primarily towards a specific, practical aim or objective. You will be provided with abstracts of academic publications. You are to rate each publication based on whether they are basic or applied research on a continuous scale from 0 to 1. A score of 0 means the research is completely basic. A score of 1 means the research is completely applied. A score of 0.5 means the research has equal basic and applied elements. 
\end{mdframed} 

\noindent $^{*}$ Prompt 5 is written in the Methods section.

\medskip

\clearpage
\subsection*{Appendix C: OLS regression results}\label{ols_regression_results}

\begin{table}[htbp]
\centering
\caption{OLS regression results on the effect of basic/applied scientists' inclusion on citation impact.}
\label{tab:ols_results}
\begin{tabular}{lcccc}
\toprule
Variable & Coefficient & Std. Error & 95\% CI & p-value \\
\midrule
Constant                          & 6.3013*** & 0.323 & [5.669, 6.933] & 0.000 \\
Basic author in                   & 0.0540*** & 0.003 & [0.049, 0.059] & 0.000 \\
Quasi Basic author in             & 0.0291*** & 0.002 & [0.025, 0.033] & 0.000 \\
Intermediate author in            & 0.0238*** & 0.002 & [0.020, 0.028] & 0.000 \\
Quasi Applied author in           & 0.0155*** & 0.002 & [0.011, 0.019] & 0.000 \\
Applied author in                 & 0.0237*** & 0.002 & [0.019, 0.029] & 0.000 \\
Team size                         & 0.0004*** & 0.000 & [0.000, 0.001] & 0.000 \\
log(team prior citations mean)         & 0.1292*** & 0.002 & [0.126, 0.132] & 0.000 \\
Established author career ages max  & -0.0021***& 0.000 & [-0.002, -0.002] & 0.000 \\
Application score                 & 0.0146**  & 0.004 & [0.006, 0.023] & 0.001 \\
Publication year                  & -0.0030***& 0.000 & [-0.003, -0.003] & 0.000 \\
Team career ages mean             & -0.0095***& 0.000 & [-0.010, -0.009] & 0.000 \\
Team expertise diversity          & -0.0089*  & 0.003 & [-0.016, -0.002] & 0.011 \\
AuthorID of First Author & \multicolumn{4}{c}{Controlled} \\
\midrule
No. Observations & \multicolumn{4}{c}{297,613} \\
No. clusters         & \multicolumn{4}{c}{282,881} \\
AIC              & \multicolumn{4}{c}{368,400} \\
R$^{2}$          & \multicolumn{4}{c}{0.127} \\
Adj. R$^{2}$     & \multicolumn{4}{c}{0.127} \\
\bottomrule
\end{tabular}
\begin{tablenotes}
\small
\item Notes: Standard errors are clustered at the first-author level. 
\item * $p<0.05$, ** $p<0.01$, *** $p<0.001$.
\end{tablenotes}
\end{table}

\end{appendices}




\end{document}